# Machine learning-based mass density model for hard magnetic 14:2:1 phases using chemical composition-based features

Anoop Kini*, Amit Kumar Choudhary, Dominic Hohs, Andreas Jansche, Hermann Baumgartl, Ricardo Büttner, Timo Bernthaler, Dagmar Goll, and Gerhard Schneider

Aalen University,

Materials Research Institute,

Beethovenstraße 1,

73430 Aalen, Germany

*Corresponding author: anoop.kini@hs-aalen.de , anoopkini@gmail.com
+49 7361 576-1620

**Abstract**

The $Fe_{14}Nd_2B$-based permanent magnets are technologically sought-after for energy conversion due to their unparalleled high energy product (520 kJ/m$^3$). For such 14:2:1 phases of different compositions, determining the magnetization from the measured magnetic moment is often bottlenecked by lack of mass density. We present a 'machine learning' (ML) mass density model for 14:2:1 phases using chemical composition-based features (representing 33 elements) and optionally lattice parameter (*LP*) features. The datasets for training and testing contain 190 phases (177 compositionally different) with their literature reported densities and *LP*. With an ML model with merely compositional features, we achieved a low mean-absolute-error of 0.51% on an unseen test-dataset.

*Keywords:* Machine learning, energy conversion, mass density, chemical composition, lattice parameters

# 1. Introduction

The 14:2:1 type permanent magnet ($TM_{14}RE_2B$; TM = transition metal, RE = rare earth, B = Boron) is an established technological material class, promising to meet the current demands for energy-conversion applications like electric automobiles [1,2], wind turbines [2], high-power motors and generators [3]. From an application viewpoint, the magnetization of the 14:2:1 phases is an important property [4–6]. Determining magnetization [7–9] of phases from magnetic moment is often not possible even if its magnetic moment gets reported, due to the absence of accurate mass density for that 14:2:1 phase.

Experimental density measurement of a 14:2:1 phase via Archimedes approach necessitates that the material to be free from any fine-sized internal porosity and other phases. Otherwise, such internal porosity and other phases get accounted for in the volume measurement, leading to an erroneous mass density measurement of a 14:2:1 phase. This approach is also effort and cost-intensive as it involves the manufacturing of a pure single-phase material.

Herbst et al. [10] calculated X-ray-based mass density for 14:2:1 phases. Apart from requiring chemical composition, the calculation strictly necessitates measured lattice parameters. They indicated that each unit cell contains exactly 4 formula units or 68 atoms [10] (containing TM sites, RE sites as well as B sites).

Unlike the density of the 14:2:1 phases, the theoretical density of multi-phased sintered magnets get reported often [11,12] (also containing grain boundary phase, oxide phase, porosity and other intermetallic phases). Empirical models get employed to arrive at 14:2:1 phase density. However, such empirical models can only be applied if the precise volume fraction and types of all phases in the microstructure are known.

The empirical model development has recently been driven using machine learning (ML) based approaches for different property predictions. Park et al. [13] illustrated an ML property model in $Fe_{14}Nd_2B$ magnets, specifically for coercivity ($\mu_0H_c$) and maximum magnetic energy product $(BH)_{max}$. Artificial neural network (ANN) and support vector regression (SVR) models were attempted ($R^2$ scores of 0.895 and 0.871). The model inputs were micromagnetic simulation data [13]. Möller et al. [14] reported anisotropy constant, $K_1$, for permanent magnets using an ML approach, however, confined to 12:1 type permanent magnets. There are no reports yet on ML-based empirical models that focus on predicting the mass density of 14:2:1 phases.

Here, we present a machine learning-based empirical model for the mass density for the magnetic 14:2:1 phase family, trained using compositional-based features. Lattice parameter features have optionally been used. The model training has been performed with a dataset containing 190 phases (124 for training and 66 for testing; a total of 177 compositionally unique). We achieved a low mean absolute error (MAE = 0.51 %) for prediction of mass density on the test set. The result is independent of presence or absence of lattice parameters.

## 2. Methods

### 2.1 Data collection for database development

A database of 190 magnetic 14:2:1 phases represented by their chemical composition was constructed. The density reported in the literature was entered for each 14:2:1 composition. When lattice parameters (*a* and *c*) were available in the literature, they were also entered into the database. For model training, 124 data entries were used, while the remaining 66 reserved for model testing. Note that the database contained 177 compositionally unique individual phases (124 unique for training and 53 unique for testing).

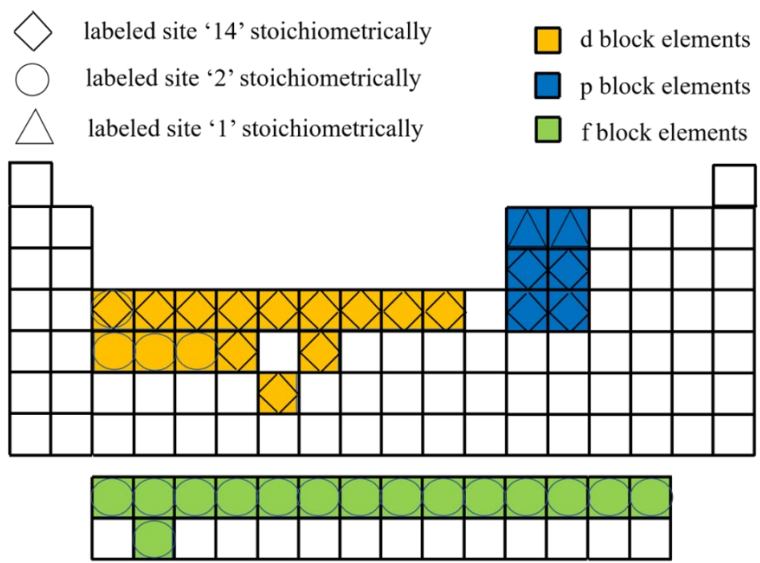

*Fig. 1: Periodic table representation of chemical elements participating in 14:2:1 permanent magnetic phases reported in literature [11,15]. These elements belong to 'd', 'p', or 'f' block denoted with orange, blue, or green colors. An element occupies one of the three stoichiometrically labeled '14', or '2', or '1' types of sites in the unit cell (represented by a diamond, circle, or triangle shape).*

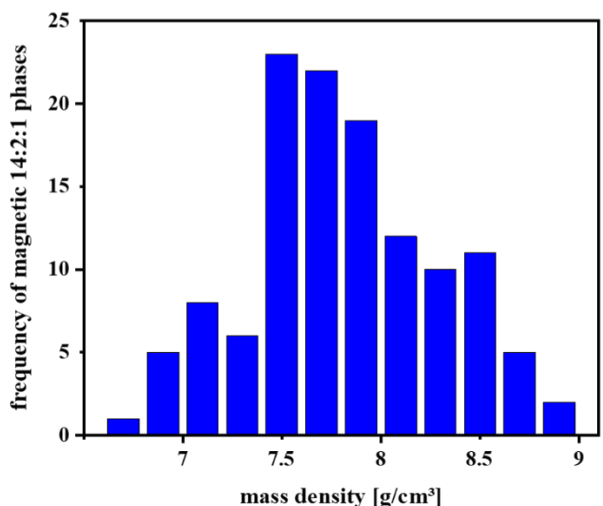

*Fig 2. Frequency distribution of mass density of 14:2:1 magnetic phases in the database according to literature reports.*

Fig. 1 shows a periodic table representation of chemical elements participating in reported 14:2:1 phases. The elements belong to 'd', 'p', or 'f' block. These elements occupy one of the three types of sites within the unit cell, labeled stoichiometrically to be '14', '2', or '1'. The frequency distribution of density for 14:2:1 phases from literature are represented as a histogram in Fig. 2.

## 2.2 Data processing

Each data entry in the database corresponds to a 14:2:1 phase represented by a chemical formula. The elemental compositional features (*C*) were extracted from the chemical formula and appended to each data entry. The *C* feature set contained 33 features, each of which represents an element described in Fig. 1. For instance, in a $Fe_{14}Nd_2B_1$ phase, the *C* features for Fe is 14, for Nd is 2, and for B is 1. The *C* features representing the remaining 30 elements were zero.

Additional composition-based features were defined by multiplying the composition (*C*) and atomic mass (*AM*) of an element in the 14:2:1 phase. The feature set has been referred to as '*C + AM*'. For instance, in a $Fe_{14}Nd_2B_1$ phase, such features for Fe, Nd, and B were 781.9, 288.48, and 10.8, while remaining 30 elements were populated with zero.

## 2.3 Machine learning models

Different supervised ML algorithms were employed for identifying regression correlation between input feature sets and the mass density. A set of 4 algorithms were screened, namely Voting Regressor (VR), Support Vector Regressor (SVR), Random Forest Regressor (RF), and Linear Regression (LR). For VR, the base regressors were SVR and RF.

The two cross-validation schemes utilized were *k*-fold [16] (*k* = 5) and leave one out cross-validation (LOOCV) [17]. The better of the two cross-validation schemes, based on mean-absolute-error (*MAE*) and regression coefficient $R^2$, was chosen for model testing.

$$MAE = \frac{100}{n} \sum_{t=1}^{n} \frac{A_t - P_t}{A_t}$$

$$R^2 = 1 - \frac{\Sigma(A_t - P_t)^2}{\Sigma(A_t - \bar{P}_t)^2}$$

Here, $A_t$ is the actual value, $P_t$ is the predicted value, and $\bar{P}_t$ is average of the predictions across the dataset containing $n$ data points.

## 3. Results

### 3.1. ML model training and cross-validation

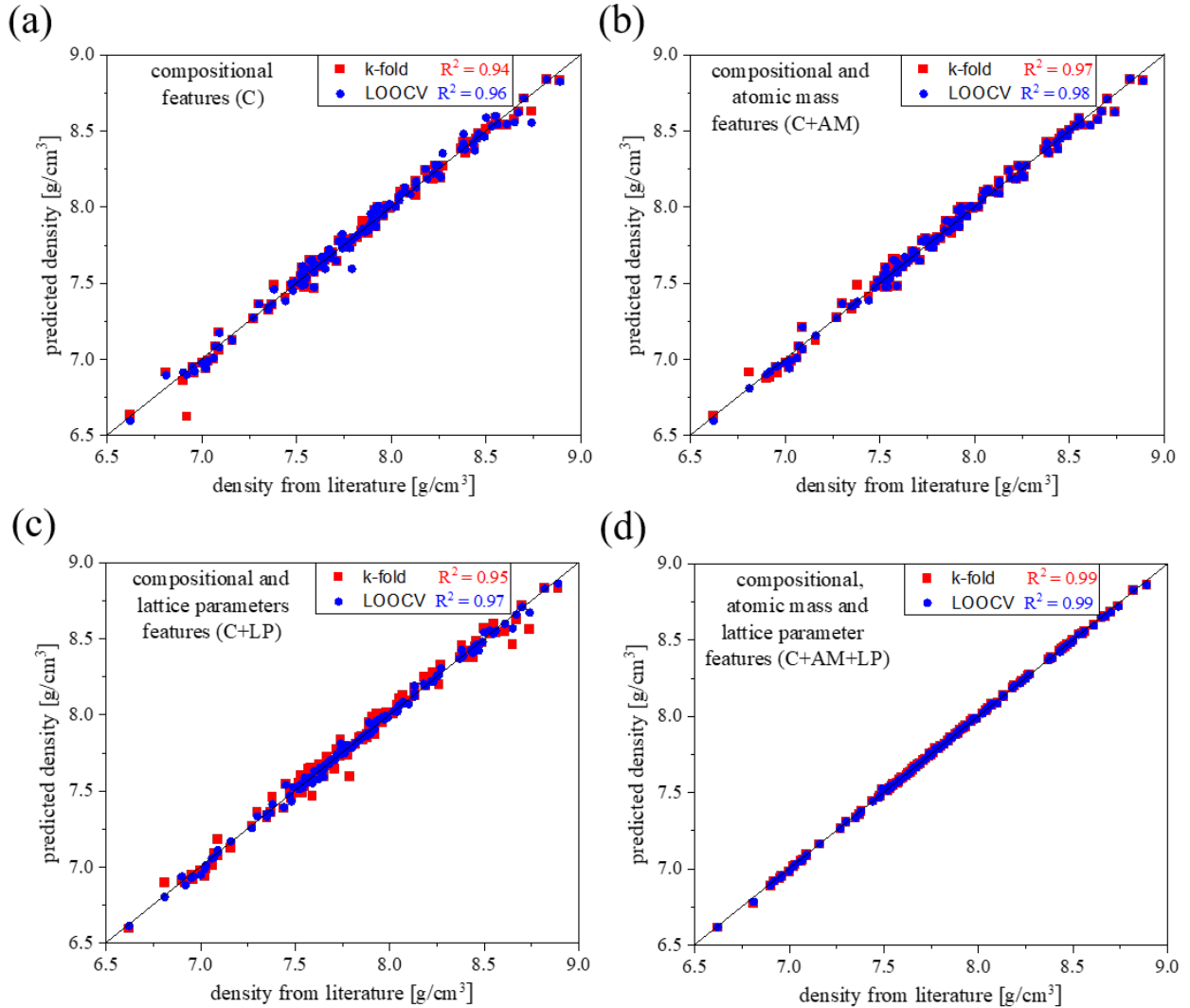

*Fig. 3: Results of machine learning (ML) model on the training set. The training set contained 124 data entries corresponding to 14:2:1 phases. The leave-one-out (LOOCV) and k-fold (k = 5) validation schemes are compared for each feature set or their combination, as denoted by each sub-figure. (a) composition (C) feature set. (b) Composition and atomic mass (C+AM) features. (c) Composition and lattice parameters (C+LP) feature set. (d) Composition, atomic mass, and lattice parameters (C+AM+LP) feature sets.*

The linear regression (LR) model performs better than other regression models on the dataset, based on *MAE* or $R^2$ as metrics. Fig. 3(a) shows the LR model results when trained with compositional features (*C*). The fit improves further if composition along with atomic mass are collectively used (*C* + *AM*) for model training, as shown in Fig. 3(b). This obtained fit is similar to that obtained with *C* + *LP* (Fig. 3(c)), i.e. combination of composition and lattice parameter feature sets. There is further marginal improvement in performance, if the model is trained with all the three feature sets, i.e. *C* + *AM* + *LP*, as displayed in Fig. 3(d).

Irrespective of the feature combinations, the LOOCV [18] cross validation scheme has been found to achieve a better fit than *k*-fold scheme ($k = 5$). The LR model confirms low underfitting. Its good generalization capability without overfitting has been elucidated next.

### 3.2. ML model evaluation on unseen test dataset

The model testing on an unseen dataset (displayed in Fig. 4), with the *C* feature set, achieves a *MAE* of 0.51 % and an $R^2$ of 0.95. With the *C*+*AM* features, *MAE* of 0.50 % and an $R^2$ of 0.97 can be secured. The performance for *C*+*LP* feature sets, resulted in *MAE* of 0.22 % and an $R^2$ of 0.98. Upon utilizing *C*+AM+*LP* feature sets the model achieves *MAE* of 0.18 % and $R^2$ of 0.99.

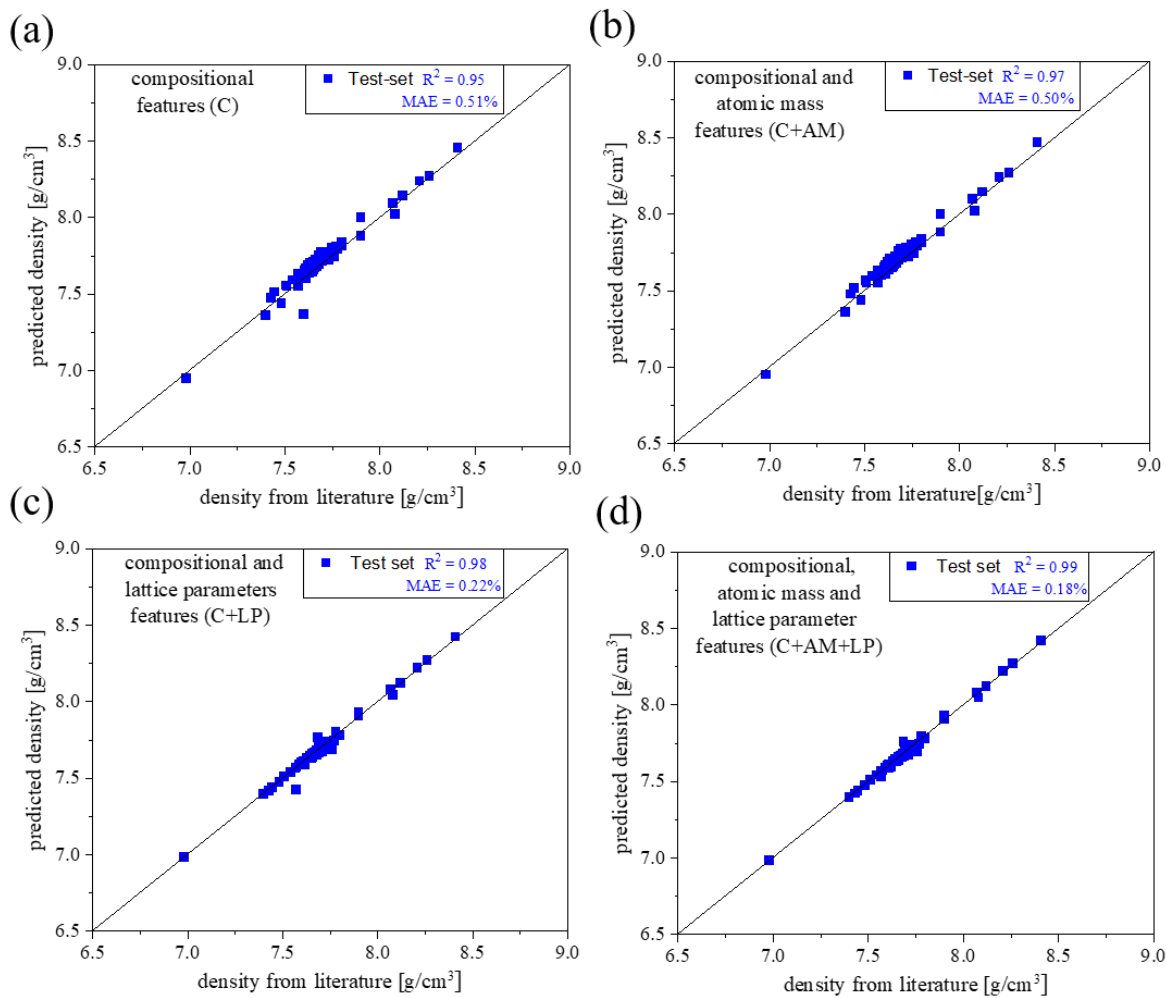

*Fig. 4: Predictions with the trained ML-based model (with LOOCV validation scheme) on an unseen test set with 66 data entries. Each sub-figure denotes a feature set or their combination as the models was trained, indicated by subfigure labels. (a) Composition (C) feature set. (b) Composition and atomic mass (C+AM) features. (c) Composition and lattice parameters (C+LP) feature sets. (d) Composition and atomic mass, as well as lattice parameters (C+AM+LP) feature sets.*

The classification of the test-set based on number of chemical elements in the phase such as ternary (3 elements), quaternary (4 elements) and quinary (5 elements). Tables 1 summarizes the performance, in terms of *MAE* for each of these classes. Although particularly for ternary phases, a few alike compositions exist in the test set, yet these are from independent literature sources with source-specific values.

| mean absolute error (MAE in %) | ternary | quaternary | quinary |
|---|---|---|---|
| C | 0.42 | 0.48 | 0.62 |
| C+AM | 0.44 | 0.42 | 0.65 |
| C+LP | 0.40 | 0.08 | 0.31 |
| C+AM+LP | 0.27 | 0.07 | 0.28 |

*Table 1: Mean absolute error (MAE) for ternary, quaternary and quinary phases associated with different feature combinations in the test-set.*

## 4. Discussion

### 4.1. Effect of composition-based features and lattice parameter features on model performance

The compositional feature set, *C*, obtained directly from the literature based chemical formula of the 14:2:1 phase, already achieves good predictive capability on unseen test dataset (*MAE* error of 0.51 %). In addition to compositional features for learning, utilizing atomic mass which are known constants, i.e. *C+AM* features, resulted in a marginally better performance (*MAE* error of 0.50 %). This shows that composition-based features, readily available from literature, to be adequate to describe mass density.

This observation reveals that the lattice parameter (*LP*) feature set, which was previously essential for calculating the X-ray density of 14:2:1 phases, is not essential for the present machine learning approach. The *LP* as a stand-alone feature set for model training resulted in a high *MAE* error of 4.1 %. Such *MAE* error is not acceptable for determining magnetization from magnetic moment data.

Note that if composition features, in addition to lattice parameters are considered, i.e. *C+LP* feature sets, it leads to a MAE of 0.22 %. The ML model performance, specifically with *C+AM+LP* features, indicates a further performance improvement and more closely learn to empirically represent density.

## 4.2. Remarks on linearity of trained ML model and predictions for compositionally new 14:2:1 phases

Linear models have previously been adequate for describing physical properties like lattice parameters of the unit cell. Some instances are perovskite systems like $ABX_3$ [19,20] or $A_2B_2X_6$ [21,22] (A: large cation with different valence, B: transition metal, and X: oxides and halides). Likewise, linear model for lattice parameters have also been used in $NiFe_2O_4$ system [23]. However, such class of multi-linear models [24] for spinel structure were used as a benchmark for comparing gaussian process regression model in a recent work by Zhang and Xu [25].

The trained ML linear model facilitates density predictions for 14:2:1 phases comprising combinations of 33 different elements (represented pictorially in Fig. 1). The equations representing the linear model for density prediction with the compositionally derived *C+AM* feature set and with *C+AM+LP* feature sets are described in the appendix.

In this 33-dimensional space, specifically for *C+AM* feature set, the mathematical complexity of the linear model is expected to be similar to the physical complexity to represent mass density. Therefore, in this 33-dimensional space, proposing a linear model, without *LP* features, and with low prediction error is not trivial. The current linear ML model, with only composition-based features and without *LP* features, achieves a prediction error of only 0.51 % *MAE* on the test-set.

## 4.3. Physical basis for the low dependence on lattice parameters (LP) in predicting the mass density of 14:2:1 phases

The TM sites occupy the majority (14 of the 17 atoms in a formula unit). Therefore, the TM sites have a much greater influence on the lattice parameters than the other two sites combined. An inspection of the entire dataset suggests that the TM sites are mainly occupied by Fe or Co. These two elements occupy at least 70 % of such sites regardless of the composition of a phase. The atomic radii of Fe or Co are nearly equal, 1.40 Å and 1.37 Å [26], which explains the lack of dependence of the model on the lattice parameters.

The substitution amount of other elements, for instance, p-block or d-block elements (Si, Ge, Ti, and Nb), is often less than 10 % [27–30]. Such low substitution fractions do not lead to a significant deviation of lattice parameters. Ni and Mn are exceptions and can substitute to a greater extent [31–33] of about 20-40 %, yet their atomic radii (1.35 Å and 1.47 Å [26]) are within only 5 % deviation from that of Fe, which does not lead to a significant deviation of the lattice parameters. The empirical ML model presented here numerically accounts for such behavior of the lattice parameters and makes generalized density predictions across 14:2:1 phases. Such a generalization was previously not possible.

The current ML models can serve as a rapid means for mass density prediction of compositionally novel 14:2:1 phases [4,6]. The key contribution of the ML model has been in demonstrating that even with only composition-based features, which are readily available from the chemical formula of the reported 14:2:1 phases, accurate predictions of density can be made. This approach can overcome the challenge of manufacturing a pure-single phase material of 14:2:1 permanent magnetic phase and also overcome the need for x-ray characterization to calculate the density using lattice parameters.

The generalized ML model for mass density prediction permits magnetization determination from measured magnetic moments for candidate 14:2:1 phases for energy conversion applications. The present work also invites opportunities for machine learning models for magnetic properties of 14:2:1 phases, using only composition-based features.

## Conclusions

1. We have developed a supervised machine learning (ML) model for predicting the mass density of magnetic 14:2:1 phases (with combinations of 33 chemical elements). The model was trained using 124 data entries corresponding to individual 14:2:1 phases with different chemical compositions. For training the ML model, the density values reported in the literature, chemical composition, and atomic masses of the contained elements were used. Lattice parameters (LP) were additionally used for training, and its effect on density prediction has also been studied.

2. The trained ML model, especially the linear regression in combination with LOOCV cross-validation, achieves a low mean absolute error (*MAE*) of less than 0.51 % and an $R^2$ greater than 0.95 when using composition-based features in the test set. The role of the *LP* feature set is not significant for generalized predictive ability.

3. The fact that the "14" sites in the lattice of the 14:2:1 unit cell are occupied by either Fe or Co by more than 70%, explains the low influence of the lattice parameters on the mass density predictions.

The present ML model allows rapid prediction of the mass density for permanent magnetic 14:2:1 phases, used popularly for energy conversion, using merely chemical composition based data. Knowledge of the mass density is relevant for determining the saturation magnetization of a 14:2:1 phase from its measured magnetic moment. The present work also opens-up avenues for developing machine learning models for magnetic properties of 14:2:1 phases, using composition-based features.

## Acknowledgements

This work was performed within the scope of the MEMORI project. The MEMORI project was made possible by funding from the Carl Zeiss Foundation. The authors gratefully acknowledge discussions with Thomas Groß and Tvrtko Grubesa from the Magnets Research Group at Aalen University.

## Appendix

The ML linear model with composition-based feature set, specifically $C+AM$ features, which are readily available in literature, has been represented using equation (1). Here, for an element 'A', the multiplication product of $C_A$ and $AM_A$, like described in section 2.2, are represented using ('partial mass'). Note that the lattice parameters feature set are not used. The weights associated with each feature has been learnt after model training.

$$mass\ density\ (in\ g/cm^3) =$$

$(0.459685988 \times PM_{Gd}) - (0.0635140839 \times PM_B) + (0.0597190135 \times PM_{Co}) + (0.195350925 \times PM_{La}) + (0.313482268 \times PM_{Nd}) + (0.292620611 \times PM_{Pr}) + (0.379187835 \times PM_{Sm}) + (0.514400205 \times PM_{Tb}) - (0.0102280072 \times PM_Y) + (0.00107446387 \times PM_{Fe}) + (0.557144297 \times PM_{Dy}) + (0.639230559 \times PM_{Er}) + (0.586626132 \times PM_{Ho}) + (0.748730789 \times PM_{Lu}) + (0.932330773 \times PM_{Th}) + (0.689745457 \times PM_{Tm}) + (0.0635140839 \times PM_C) + (0.385532139 \times PM_{Ce}) - (0.21230264 \times PM_{Al}) - (0.332445864 \times PM_{Zr}) + (0.0167135759 \times PM_{Cu}) + (0.0566577464 \times PM_{Ga}) + (0.102598472 \times PM_{Ge}) - (0.0260357843 \times PM_{Mn}) + (0.0584528744 \times PM_{Ni}) - (0.107308321 \times PM_{Si}) + (0.037304529 \times PM_{Cr}) - (0.104485726 \times PM_{Sc}) + (2.08166817E - 17 \times PM_{Mo}) + (0.265824833 \times PM_{Nb}) + (0.304619184 \times PM_{Ru}) - (0.0797266208 \times PM_{Ti}) - (0.0346932378 \times PM_V) - 7.021582$ (1)

The ML linear model developed with collective $C$, $AM$ and $LP$ feature sets are shown next in equation (2). The LP features are represented using $a$ and $c$. The combination of $C+AM$ features, like in equation 1, has been represented using $PM$. The subscripts here also refer to elements.

$$mass\ density\ (in\ g/cm^3) =$$
$-(18.9309216 * a) - (6.43821730 * c) - (3.66628727e - 05 * C_{Sc}) - (2.21720852e - 05 * C_{Ti}) - (1.48170020e - 05 * C_V) - (1.26132937e - 05 * C_{Cr}) - (2.79679440e - 06 * C_{Mn})$

$$\begin{aligned}
&-(8.15233511e-07 * C_{Fe}) + (5.68413616e-06 * C_{Co}) + (4.63373805e-06 * C_{Ni}) \\
&+(1.34915341e-05 * C_{Cu}) - (2.85926637e-04 * C_{Al}) - (2.69462162e-04 * C_{Si}) \\
&+(1.97808557e-05 * C_{Ga}) + (2.26738100e-05 * C_{Ge}) + (3.21914872e-05 * C_{Nb}) \\
&+(3.20774848e-05 * C_{Mo}) + (2.65399038e-05 * C_{Re}) + (3.09641608e-05 * C_{Ru}) \\
&+(3.21603031e-05 * C_{La}) + (3.18687734e-05 * C_{Ce}) + (3.19606808e-05 * C_{Pr}) \\
&+(3.16267367e-05 * C_{Nd}) + (3.11950882e-05 * C_{Sm}) + (3.03505161e-05 * C_{Gd}) \\
&+(3.04023626e-05 * C_{Tb}) + (2.98604904e-05 * C_{Dy}) + (2.95821068e-05 * C_{Ho}) \\
&+(2.93176366e-05 * C_{Tm}) + (2.88661317e-05 * C_{Lu}) + (2.37290838e-05 * C_{Th}) \\
&+(3.20840870e-05 * C_{Y}) + (3.47937965e-05 * C_{Zr}) - (3.79784870e-05 * C_{B}) \\
&+(3.79784869e-05 * C_{C}) - (1.64821334e-03 * PM_{Sc}) - (1.06145358e-03 * PM_{Ti}) \\
&-(7.54805634e-04 * PM_{V}) - (6.55841128e-04 * PM_{Cr}) - (1.53650128e-04 * PM_{Mn}) \\
&-(4.55282037e-05 * PM_{Fe}) + (3.34982191e-04 * PM_{Co}) + (2.71960907e-04 * PM_{Ni}) \\
&+(8.57333629e-04 * PM_{Cu}) - (7.71487249e-03 * PM_{Al}) - (7.56804680e-03 * PM_{Si}) \\
&+(1.37918111e-03 * PM_{Ga}) + (1.64668599e-03 * PM_{Ge}) + (2.99078205e-03 * PM_{Nb}) \\
&+(3.07751378e-03 * PM_{Mo}) + (4.94191736e-03 * PM_{Re}) + (3.12954621e-03 * PM_{Ru}) \\
&+(4.46726008e-03 * PM_{La}) + (4.46531035e-03 * PM_{Ce}) + (4.50351485e-03 * PM_{Pr}) \\
&+(4.56187127e-03 * PM_{Nd}) + (4.69049378e-03 * PM_{Sm}) + (4.77261873e-03 * PM_{Gd}) \\
&+(4.83169557e-03 * PM_{Tb}) + (4.85232977e-03 * PM_{Dy}) + (4.87897686e-03 * PM_{Ho}) \\
&+(4.95274561e-03 * PM_{Tm}) + (5.05062039e-03 * PM_{Lu}) + (5.50604915e-03 * PM_{Th}) \\
&+(2.85246784e-03 * PM_{Y}) + (3.17402929e-03 * PM_{Zr}) - (4.10585419e-04 * PM_{B}) \\
&+(4.56153907e-04 * PM_{C}) + 30.838
\end{aligned} \qquad (2)$$